\documentclass[twocolumn,aps,superscriptaddress,showpacs,nofootinbib]{revtex4}
\usepackage{hyperref}
\usepackage{footnote}
\usepackage{amssymb}
\usepackage{amsmath}
\usepackage{graphicx}
\usepackage[normalem]{ulem}
\usepackage{url}
\usepackage{csquotes}
\usepackage{color}

\begin{document}
\title{Sensitivity of the evaporation residue observables to the symmetry energy}

\author{S. Mallik}
\affiliation{Physics Group, Variable Energy Cyclotron Centre, 1/AF Bidhan Nagar, Kolkata 700064, India}

\author{G. Chaudhuri}
\affiliation{Physics Group, Variable Energy Cyclotron Centre, 1/AF Bidhan Nagar, Kolkata 700064, India}
\affiliation{Homi Bhabha National Institute, Training School Complex, Anushakti Nagar, Mumbai 400085, India}
\author{F. Gulminelli}
\affiliation{LPC Caen IN2P3-CNRS/EnsiCaen et Universite, Caen, France}

\begin{abstract}
The static properties of the heaviest residue and unbound particles produced  in central $^{64,58}Ni$ on $^{64,58}Ni$   collisions at 50 MeV/nucleon are
predicted within the BUU transport model, in order to explore the sensitivity of those observables to the density dependence of the symmetry energy.
We include fluctuations in the collision integral and use a meta-modelling for the mean-field which allows an independent variation of the different  empirical parameters
of the equation of state. We find that the isospin ratio of pre-equilibrium particles is a good
estimator of the stiffness of the symmetry energy, in agreement with previous works. In addition to that,
whatever be the functional form of the equation of state, we show that a higher symmetry energy at subsaturation densities leads to an increased size and isotopic ratio for the heaviest residue. This is understood in terms of energy sharing between the pre-equilibrium particles and the (quasi)fused system. The combination of the two observables might be an interesting tool to constrain the different density dependence below and above saturation, which is linked to the relatively poorly known parameter $K_{sym}$.
\end{abstract}

\pacs{25.70Mn, 25.70Pq,
64.10.+h, 
64.60.-i, 
24.10.Pa 
}
\maketitle
\section{Introduction}
The   equation of state (EoS) is one of the fundamental properties governing nuclear systems, and its theorietical as well as
experimental determination is an extremely lively issue in modern nuclear physics and astrophysics \cite{oertel16}.
The biggest uncertainties concern the so called symmetry energy, which describes the density behavior of strongly asymmetric matter,
and which is extremely important for the understanding of a large variety of astrophysical phenomena involving compact stars \cite{epja50,Lattimer16}.\\
\indent
From the historical point of view, the first attempts in constraining the symmetry energy at densities different from the saturation density of symmetric nuclear matter
used experimental observables from intermediate energy heavy ion collisions \cite{tsang2004}.
Since that time, different experimental probes from nuclear structure have contributed to better constrain the symmetry energy behavior \cite{tsang_rep,horowitz2014}, and clear correlations between the different symmetry energy empirical parameters were convincingly extracted from this data compilation
 ~\cite{Kortelainen2012,Danielewicz2014,Trippa2008,Colo2014,Warda2009,Centelles2010,Mondal2016}.
Still, heavy ion collisions are the only laboratory condition where nuclear matter is effectively compressed, and in this sense they can be considered as a privileged tool for studying the behavior of symmetry energy, especially in the super-saturation regime \cite{russotto2016}.\\
\indent
The extraction of symmetry energy from HI collisions requires the comparison of an isospin-sensitive observable to the predictions of transport calculations.
However, most of these calculations use a very simplistic monotropic functional form for the symmetry energy  \cite{epja50,tsang_rep}, $e_{sym}(\rho)\propto \rho^\gamma$,
 which is not justified
in the framework of modern energy functionals. For this reason, it is not easy to quantitatively compare the constraints extracted from HI collisions with the ones obtained from structure experiments which are analyzed with Skyrme or RMF functionals, including a more complex density dependence as well as neutron-proton mass splittings.\\
\indent
In this work, we have implemented the meta-functional proposed in ref.\cite{Margueron2018a} in the improved  BUU@VECC-McGill transport code \cite{Mallik9,Mallik11}.
The parameter space of the meta-model allows to precisely reproduce a large set of non-relativistic as well as relativistic energy functionals, as well as possible novel density dependences
not yet explored in existing functionals. Moreover, the symmetry energy empirical parameters can be independently varied, allowing a sensitivity analysis of the HI observables. The  BUU@VECC-McGill transport code includes isospin dependent nucleon-nucleon cross sections and successfully compared to other transport models \cite{Zhang_code_comparison,Ono_code_comparison} as well as experimental data \cite{Mallik23}.\\
\indent
Our purpose is to guide and inspire future HI experiments and experimental analyses, by proposing  measurable observables sensitive to the different parameters of the symmetry energy.
For this first application, we concentrate on central $^{64}Ni$ on $^{64}Ni$, $^{58}Ni$ on $^{58}Ni$, and $^{58}Ni$ on $^{64}Ni$ collisions
at 50 MeV/nucleon beam energy. The choice of the system is due to the fact that these systems will be studied by the INDRA/FAZIA collaboration in an upcoming experiment
in GANIL this year \cite{indra}.\\
\indent
We concentrate on the composition of the heaviest fragment as well as the free nucleons, both at freeze-out and at asymptotic times. Results will be given for the realistic Sly5 functional \cite{Sly5}
\cite{Sly5}, and a sensitivity analysis to the different empirical parameters will be performed.
\section{Theoretical framework}
\subsection{The BUU model}
The dynamical evolution of the heavy ion collision is followed using the  BUU@VECC-McGill transport model calculation, which was extensively explained in Refs. \cite{Mallik9,Mallik11}.
The calculation  is started when two nuclei in their respective ground states approach each other with specified velocities.\\
\indent
At each time step, the local particle densities $\rho_q$ and kinetic energy densities $\tau_q$ ($q=n,p$) are defined from the one-body distribution functions as:
\begin{eqnarray}
\rho_q(\vec r, t)&=&\int d^3p f_q(\vec r, \vec p, t) \nonumber \\
\tau_q(\vec r, t)&=&\int d^3p \frac{p^2}{2 m_q}  f_q(\vec r, \vec p, t) \ , \label{eq:fq}
\end{eqnarray}
where the distribution functions $f_q$ are sampled with test particles that follow Hamilton equations of motion \cite{Mallik9,Mallik11},
and we neglect the difference between the proton and neutron bare masses, $m_q=m$.
At the initial time, the ground state densities of the projectile (target)  of mass number $A_P$ ( $A_T$ ) , where  $A_k=Z_k+N_k$  (k=P,T) and $Z_k$ and $N_k$ are proton  and neutron numbers,    are constructed by a variational method \cite{Mallik9} using Myers density profiles \cite{Myers}. This method was used by different authors \cite{Mallik9,Lee,Cecil}. The ground state density distribution is then sampled using a Monte-Carlo technique by choosing $N_{test}=100$ test particles  for each nucleon,  with appropriate positions and momenta.\\
\indent
In the center of mass frame, the test particles of the projectile and the target nuclei 
are boosted towards each other. Simulations are done in a 200$\times 200\times 200 fm^3$ box. At t=0 fm/c the projectile and target nuclei are centered at (100 fm,100 fm,90 fm) and (100 fm,100 fm,110 fm). The test particles of isospin $q=p,n$ move in a mean-field $U_q(\rho_p(\vec{r}),\rho_n(\vec{r}))$ and will occasionally suffer two-body collisions, with probability determined by the nucleon-nucleon scattering cross section, provided the final state of the collision is not blocked by the Pauli principle. The mean field potential $U_q$ is calculated from a meta-functional  described in the next subsection. The mean-field propagation is done using the lattice Hamiltonian method which conserves energy and momentum very accurately \cite{Lenk}. Two body collisions are calculated as in Appendix B of ref. \cite{Dasgupta_BUU1}, except that pion channels are closed, as there will not be any pion production in this energy regime.\\
\indent
To explain clustering in heavy ion reaction, one needs an event-by-event computation in transport calculation, and mean-field fluctuations should be accounted for \cite{Bauer}. To do that, we have followed the recently developed computationally efficient prescription described in Ref. \cite{Mallik10,Mallik14,Mallik18}, which leads to a correct propagation if the collision partners contain a  sufficiently large number of nucleons.   According to this prescription, the nucleon-nucleon collisions are computed  at each time step with the physical isospin dependent cross-section 
only among the $A_P+A_T$ test-particles belonging to the same event. For each event, if a collision between two test particles $i$ and $j$ is allowed, the method proposed in ref. \cite{Bauer,Mallik10} is  followed: the ($N_{test}-1$) test particles closest to $i$ in configuration space are  picked up, and the same momentum change $\Delta \vec{p}$ as ascribed to $i$ is  given to all of them. Similarly the ($N_{test}-1$) test particles closest to $j$ are  selected and these  are  ascribed the same momentum change $-\Delta \vec{p}$ suffered by $j$. As a function of time this is continued till the event is over and the same procedure is repeated for each event. We consider free cross section parameterized from experimental data. Finally to identify fragments, two test particles are considered as the part of the same cluster if the distance between them is less than or equal to $2$ fm \cite{Mallik14}.
\subsection{The EoS meta-modelling}
A meta-modelling approach for the  nucleonic equation of state was proposed in ref. \cite{Margueron2018a}. A flexible functional was proposed,
based on a polynomial expansion in density around saturation and including deviations from the parabolic isospin dependence through the kinetic term and the effective mass splitting.
The parameter space of this functional is sufficiently large to allow reproducing with good accuracy a large set of popular relativistic and non-relativistic functionals.
The polynomial expansion implies that the different empirical parameters are a-priori independent, and it is therefore possible to study the effect of an independent variation of each of them, which we do in the present work. This functional was already applied to neutron star observables \cite{Margueron2018b}, finite nuclei \cite{Chatterjee2018}, and magnetars \cite{Chatterjee2019}.

 The energy per particle of homogeneous nuclear matter at zero temperature and proton (neutron) density $\rho_p$ ($\rho_n$)  is
\begin{equation}
e(\rho,\delta)=t(\rho,\delta)+v(\rho,\delta) \, ,
\label{Total_energy_per_nucleon_EFFc}
\end{equation}
where $\rho=\rho_p+\rho_n$ and $\delta=(\rho_n-\rho_p)/\rho$.
The kinetic energy per particle at zero temperature, including an effective  momentum dependence through the definition of effective masses, is given by
\begin{eqnarray}
t^*(\rho,\delta)&=&\frac{t_0}{2}\bigg{(}\frac{\rho}{\rho_0}\bigg{)}^{2/3}\bigg{[}(1+\kappa_0\frac{\rho}{\rho_0})f_1(\delta)\nonumber\\
&+&\kappa_{sym}\frac{\rho}{\rho_0}f_2(\delta)\bigg{]}
\label{Kinetic_energy_per_nucleon_EFFc}
\end{eqnarray}
where  $t_{0}= 3\hbar^{2}/(10m)\left(3\pi^{2}/2\right)^{2/3}\rho_{0}^{2/3}$  and $\rho_0$ is the (model dependent) saturation density of symmetric nuclear matter.
The parameters  $\kappa_0$ and  $\kappa_{sym}$  are linked to the density dependence of the effective proton and neutron masses,
and   $f_1=\{(1+\delta)^{5/3}+(1-\delta)^{5/3}\}$, $f_2=\delta\{(1+\delta)^{5/3}-(1-\delta)^{5/3}\}$.\\
\indent
The expression of potential energy per particle is
\begin{eqnarray}
v(\rho,\delta)&=&\sum_{k=0}^{N}\frac{1}{k!}(v^{is}_{k}+v^{iv}_{k}\delta^2)x^{k}\nonumber\\
&+&(a^{is}+a^{iv}\delta^2)x^{N+1}\exp(-b\frac{\rho}{\rho_0}) ,
\label{ELFc_potential}
\end{eqnarray}
where $x=(\rho-\rho_0)/3\rho_0$, $\rho_0$ is the saturation density, and the last term
is a low density correction ensuring the correct limit at zero density. We take for this paper $N=4$ and $b=10ln2$. This value of $b$ leads to a good reproduction of the Sly5 functional which will be our reference model in this study. The model parameters $v_k^{is(iv)}$ can be linked with a one-to-one correspondence to the usual EoS empirical parameters, via:
\begin{eqnarray}
v^{is}_{0}&=&E_{sat}-t_0(1+\kappa_0)\nonumber\\
v^{is}_{1}&=&-t_0(2+5\kappa_0)\nonumber\\
v^{is}_{2}&=&K_{sat}-2t_0(-1+5\kappa_0)\nonumber\\
v^{is}_{3}&=&Q_{sat}-2t_0(4-5\kappa_0)\nonumber\\
v^{is}_{4}&=&Z_{sat}-8t_0(-7+5\kappa_0)
\label{Isoscalar_parameters}
\end{eqnarray}
\begin{eqnarray}
v^{iv}_{0}&=&E_{sym}-\frac{5}{9}t_0[(1+(\kappa_0+3\kappa_{sym})]\nonumber\\
v^{iv}_{1}&=&L_{sym}-\frac{5}{9}t_0[(2+5(\kappa_0+3\kappa_{sym})]\nonumber\\
v^{iv}_{2}&=&K_{sym}-\frac{10}{9}t_0[(-1+5(\kappa_0+3\kappa_{sym})]\nonumber\\
v^{iv}_{3}&=&Q_{sym}-\frac{10}{9}t_0[(4-5(\kappa_0+3\kappa_{sym})]\nonumber\\
v^{iv}_{4}&=&Z_{sym}-\frac{40}{9}t_0[(-7+5(\kappa_0+3\kappa_{sym})] \ ,
\label{Isovector_parameters}
\end{eqnarray}
where  $E_{sat}$, $K_{sat}$, $Q_{sat}$ and $Z_{sat}$ are saturation energy, incompressibility modulus,  isospin symmetric skewness and  kurtosis respectively and $E_{sym}$, $L_{sym}$, $K_{sym}$, $Q_{sym}$ and $Z_{sym}$ are symmetry energy, slope, and associated incompressibility,  skewness and  kurtosis respectively.\\
\indent
This infinite nuclear matter functional is supplemented by a finite range term which from the theoretical point of view  arises from the semi-classical $\hbar$ expansion of the non-local momentum operator \cite{Brack1985,Lenk}, corresponding to an energy density $e_{surf}=A\rho\nabla^2 \rho$. For this first application, we neglect the isospin dependence of this gradient term \cite{Chatterjee2018} and fix the coupling parameter $A=c/(2\rho_0^{5/3})$ with $c=-6.5$ MeV from ref.\cite{Lenk}.

 In terms of the one-body distribution functions used in BUU, the total energy can be written as:

\begin{eqnarray}
E_{tot}(t)&=& \int d^3 r \epsilon \left (\rho(\vec r),\delta(\vec r)\right)  \label{eq:dense}\\
&=&
\sum_{q=n,p} \int d^3 r \frac{m}{m^*\left (\rho(\vec r,t),\delta (\vec r,t)\right )} \tau_q(\vec r,t) \nonumber \\
&+&  \int d^3 r \rho(\vec r,t) v\left (\rho(\vec r,t),\delta (\vec r,t)\right ) . \label{eq:etot}
\end{eqnarray}

Here, the local kinetic energy densities aregiven by eq.(\ref{eq:fq}) and naturally deviate as a function of time from the
zero temperature initial condition. The  local effective masses are given by:

\begin{equation}
\frac{m_q}{m^*_q}=1+(\kappa_0 \pm \kappa_{sym}\delta)\frac{\rho}{\rho_{0}} \ ,
\end{equation}

and the sign +(-) refers to neutrons (protons).


The mean-field potential governing the equations of motion of test particles can be straightforwardly obtained from the energy density defined in eq.(\ref{eq:dense}) in the local density approximation,
from the general relations:
\begin{equation}
\label{Neutron_mean_field_1}
U_n(\vec r,t)=\bigg{(}\frac{\partial \epsilon}{\partial \rho_n}\bigg{)}_{\rho_p,\tau_p,\tau_n} \; ; \;
U_p(\vec r,t)=\bigg{(}\frac{\partial \epsilon}{\partial \rho_p}\bigg{)}_{\rho_n,\tau_p,\tau_n}.
\end{equation}

Substituting eq. (\ref{ELFc_potential}) in eq. (\ref{Neutron_mean_field_1})  and adding the finite range 
the potential part of neutron and proton mean fields for BUU calculation are
\begin{eqnarray}
U_{n,loc}&=&(v^{is}_{0}+v^{iv}_{0}\delta^2)+\sum_{k=1}^{4}\frac{k+1}{k!}(v^{is}_{k}+v^{iv}_{k}\delta^2)x^{k}\nonumber\\
&+&\frac{1}{3}\sum_{k=1}^{4}\frac{1}{(k-1)!}(v^{is}_{k}+v^{iv}_{k}\delta^2)x^{k-1}\nonumber\\
&+&2\delta(1-\delta)\sum_{k=1}^{4}\frac{1}{k!}v^{iv}_{k}x^{k}+\exp\{-b(1+3x)\}\nonumber\\
&\times&\bigg{[}(a^{is}+a^{iv}\delta^2)\bigg{\{}\frac{5}{3}x^4+(6-b)x^{5}\nonumber\\
&-&3bx^{6}\bigg{\}}+2\delta(1-\delta)a^{iv}x^{5}\bigg{]}+\frac{3c}{\rho_0^{2/3}}\nabla^2x
\label{Neutron_mean_field_2}
\end{eqnarray}
\begin{eqnarray}
U_{p,loc}&=&(v^{is}_{0}+v^{iv}_{0}\delta^2)+\sum_{k=1}^{4}\frac{k+1}{k!}(v^{is}_{k}+v^{iv}_{k}\delta^2)x^{k}\nonumber\\
&+&\frac{1}{3}\sum_{k=1}^{4}\frac{1}{(k-1)!}(v^{is}_{k}+v^{iv}_{k}\delta^2)x^{k-1}\nonumber\\
&-&2\delta(1+\delta)\sum_{k=1}^{4}\frac{1}{k!}v^{iv}_{k}x^{k}+\exp\{-b(1+3x)\}\nonumber\\
&\times&\bigg{[}(a^{is}+a^{iv}\delta^2)\bigg{\{}\frac{5}{3}x^4+(6-b)x^{5}\nonumber\\
&-&3bx^{6}\bigg{\}}-2\delta(1+\delta)a^{iv}x^{5}\bigg{]}+\frac{3c}{\rho_0^{2/3}}\nabla^2x \ ,
\label{Proton_mean_field_2}
\end{eqnarray}
where $x=(\rho(\vec r,t)-\rho_0)/{3\rho_0}$ and $\delta=(\rho_n(\vec r,t)-\rho_p(\vec r,t))/\rho(\vec r,t)$.

The density dependence of the effective masses induces an extra term for the mean field given by:

\begin{eqnarray}
U_q^{eff}&=&\sum_{q=n,p} \tau_q \frac{\partial}{\partial \rho_q}\left ( \frac{m_q}{m^*_q}\right ) \nonumber \\
&=&\tau_q \frac{\kappa_0+\kappa_{sym}}{\rho_0} + \tau_{q'} \frac{\kappa_0-\kappa_{sym}}{\rho_0}
\end{eqnarray}

The complete parameter set of the meta-modelling comprises the 10 EoS empirical parameters ($\rho_0$, $E_{sat}$, $K_{sat}$, $Q_{sat}$, $Z_{sat}$,
$E_{sym}$, $L_{sym}$, $K_{sym}$, $Q_{sym}$, $Z_{sym}$), the two parameters defining the density dependence of the effective mass and the proton-neutron mass splitting
($\kappa_0$, $\kappa_{sym}$), and the finite size parameter $c$. This is a very large parameter space, and for this first application we will neglect the density dependence of the effective mass and the mass splitting, which are expected to be less influential than the EoS parameters \cite{Chatterjee2018}.
Concerning these latters, only the lowest order ones are influential at the low densities studied here, and since we are interested in pinning down isospin effects we will only concentrate on an independent variation of the lowest order isospin dependent parameters, namely $E_{sym}$, $L_{sym}$, $K_{sym}$. Values of fixed parameters ($\rho_0$, $E_{sat}$, $K_{sat}$, $Q_{sat}$, $Z_{sat}$, $Q_{sym}$, $Z_{sym}$, $\kappa_0$ and $\kappa_{sym}$) and varying parameters ($E_{sym}$, $L_{sym}$, $K_{sym}$) used for the calculations described in section III.B and III.C are given in Table-\ref{table_parameter}.

\begin{table}
\begin{center}
\begin{tabular}{|p{1.45cm}|p{1.14cm}|p{1.45cm}|p{1.14cm}|p{1.45cm}|p{1.14cm}|}
\hline
\multicolumn{6}{|c|}{Fixed Parameters} \\
\hline
Parameter & Average & Parameter & Average & Parameter & Average \\
 & Value & & Value & & Value \\
 \hline
$E_{sat}$ & -16.03 & $K_{sat}$ &251 & $Q_{sat}$ & 13 \\
(MeV)& & (MeV)& & (MeV)& \\
\hline
$Z_{sat}$ &3925 & $Q_{sym}$ & 388 & $Z_{sym}$ &-5268   \\
(MeV)& & (MeV)& & (MeV)&  \\
\hline
$\rho_0$ &0.1543 & $\kappa_0$ &0.338 & $\kappa_{sym}$ &-0.002 \\
(fm$^{-3})$& & & & & \\
\hline
\end{tabular}
\begin{tabular}{|p{2.005cm}|p{2.005cm}|p{2.005cm}|p{2.005cm}|}
\hline
\multicolumn{4}{|c|}{Varying Parameters} \\
\hline
Parameter & Minimum $    $     & Average     & Maximum     \\
 & Value & Value & Value \\
 \hline
$E_{sym}$ & 26.83 & 33.30 & 38.71 \\
(MeV)& & & \\
\hline
$L_{sym}$ & 29.2 & 76.6 & 122.7 \\
(MeV)& & & \\
\hline
$K_{sym}$ & -394 & -3 & 213 \\
(MeV)& & & \\
\hline
\end{tabular}
\end{center}
\caption{Set of fixed parameters (upper part) and varying parameters (lower part) used for the calculations described in section III.B and III.C}
\label{table_parameter}
\end{table}

\section{Results}
\subsection{Freeze-out time and asymptotic time}\label{sec:times}

\begin{figure}[b]
\begin{center}
\includegraphics[width=0.7\columnwidth]{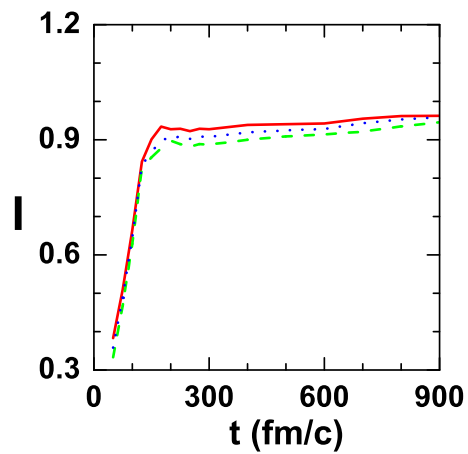}
\caption{Variation of isotropy of momentum distribution ($I$) with time for $^{58}Ni$ on $^{58}Ni$ (green dashed line), $^{58}Ni$ on $^{64}Ni$ (blue dotted line), $^{64}Ni$ on $^{64}Ni$ (red solid line) at projectile beam energy 50 MeV/nucleon.}
\label{Isotropy}
\end{center}
\end{figure}

We first determine meaningful times to be considered for further analyses. For this purpose,  we consider the Sly5 interaction.
Central $b=0$  $^{64}Ni$ on $^{64}Ni$, $^{58}Ni$ on $^{58}Ni$, and $^{58}Ni$ on $^{64}Ni$ collisions
at 50 MeV/nucleon beam  energy have been simulated up to 900 fm/c.
The freeze out time can be identified from the behavior of   the isotropy ratio as a function of time, as shown in Figure \ref{Isotropy}.  The isotropy ratio is defined as
\begin{equation}
I=\frac{\langle( p_x-\langle p_x\rangle)^2\rangle + \langle( p_y-\langle p_y\rangle)^2\rangle}{2\langle( p_z-\langle p_z\rangle)^2\rangle},
\end{equation}
where the average is taken over the test-particles belonging to the heaviest residue and $z$ is the beam axis.
 We can see that full equilibrium of the momentum distribution is never completely reached, but the collisional dynamics which tends to randomize the momenta of the nucleons is over at at $t_{FO}=150$ fm/c, and this freeze out time does not  change for the three systems considered.

\begin{figure}[t]
\begin{center}
\includegraphics[width=\columnwidth]{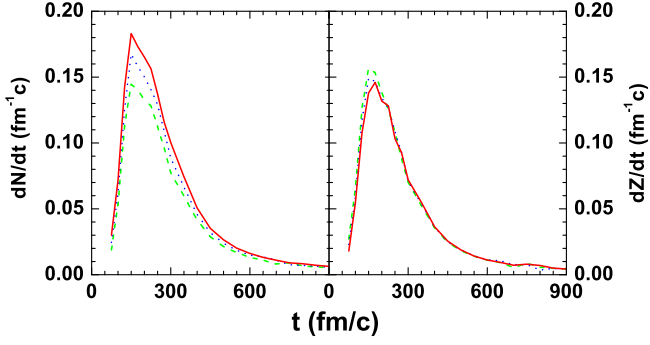}
\caption{Variation of neutron(left part) and proton (right part) emission rate as a function of time for central  for $^{58}Ni$ on $^{58}Ni$ (green dashed line), $^{58}Ni$ on $^{64}Ni$ (blue dotted line), $^{64}Ni$ on $^{64}Ni$ (red solid line) reaction at 50 MeV/nucleon.  }
\label{free_particles_time}
\end{center}
\end{figure}
\begin{figure}[b]
\begin{center}
\includegraphics[width=0.7\columnwidth]{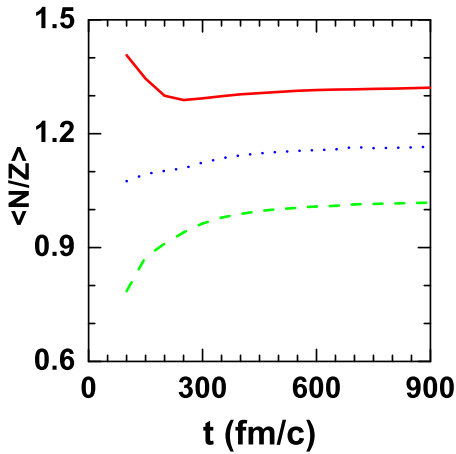}
\caption{Variation of the $\langle N \rangle/\langle Z \rangle$ of light particles emitted as a function of time, for $^{58}Ni$ on $^{58}Ni$ (green dashed line), $^{58}Ni$ on $^{64}Ni$ (blue dotted line), $^{64}Ni$ on $^{64}Ni$ (red solid line) reaction at 50 MeV/nucleon.}
\label{N_by_Z_time_depndence}
\end{center}
\end{figure}
This choice of freeze-out time is confirmed by inspection of Figure \ref{free_particles_time} which displays the free proton and neutron emission rates as a function of time for the three systems. The presence of a peak in the emission rate
is a clear indication of change of emission mechanism for particle production, from pre-equilibrium fast emission at early times to nucleon evaporation at later times \cite{li_report}. From these observations, we will keep $t=150$ fm/c  as freeze-out time.\\
\indent
It is customary in HI transport calculations to stop the dynamical evolution at the freeze-out time and couple the calculation to a statistical decay code or afterburner\cite{li_report}. Such a procedure allows calculating realistic multiplicities for the light particles which would not be correctly bound in the mean-field approximation implicit in the transport equation.
However, the precise choice of the coupling time and of the algorithm used for the calculation of excitation energy is delicate, because the final yields depend on these unconstrained parameters. Moreover, there might be some conceptual inconsistency between the mean-field model used for the dynamical evolution and the level density and mass model used for the secondary decay. This can create ambiguities in a study like the present one aimed at exploring the sensitivity to the symmetry energy functional.
For these reasons, we do not couple the dynamical code to an afterburner and rather continue the evolution to an asymptotic time where the evaporation dynamics is essentially over. The drawback of this procedure is that we will not be able to give realistic predictions for the light particles yields (only free protons and neutrons are evaporated), but we believe that global collective variables such as the global average N/Z ratio of the emitted particles, and the size and charge of the evaporation residue, will be reasonably trustable for a comparison with experimental data.\\
\indent
Since we want to study the sensitivity to the isospin part of the equation of state, we fix our asymptotic time on the saturation of the N/Z ratio of the emitted light particles. This is shown in Figure \ref{N_by_Z_time_depndence} for the three systems studied in this paper. We can see that a saturation of this ratio is achieved starting from $t$=500 fm/c, and again this time is seen to be independent of the entrance channel.\\
\begin{figure}[t]
\begin{center}
\includegraphics[width=\columnwidth]{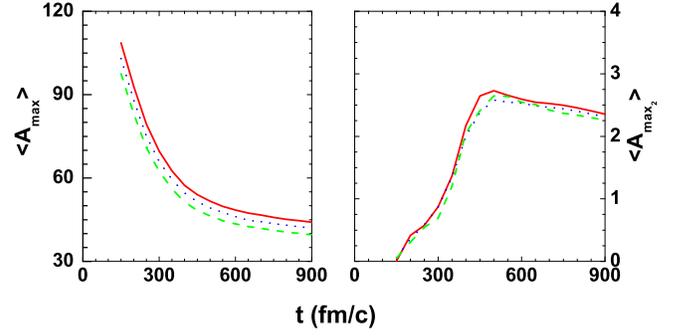}
\caption{Variation of largest (left)  and second largest (right) cluster mass  with time for $^{58}Ni$ on $^{58}Ni$ (green dashed line), $^{58}Ni$ on $^{64}Ni$ (blue dotted line), $^{64}Ni$ on $^{64}Ni$ (red solid line) at projectile beam energy 50 MeV/nucleon.}
\label{Largest_and_second_largest_vs_time}
\end{center}
\end{figure}
\begin{table}
\begin{center}
\begin{tabular}{|c|c|c|c|c|c|}
\hline
& &\multicolumn{2}{|c|}{Result at} & \multicolumn{2}{c|}{Result at} \\
& &\multicolumn{2}{|c|}{freeze-out time} & \multicolumn{2}{c|}{asymptotic time} \\
Observable & Reaction &\multicolumn{2}{|c|}{($t=150$ fm/c)} & \multicolumn{2}{c|}{($t=500$ fm/c)} \\

\cline{3-6}
 &  & Average & Standard & Average & Standard \\
 &  &  & deviation &  & deviation \\
\hline
& $^{58}$Ni+$^{58}$Ni & 0.878 & 0.077 & 1.001 & 0.023 \\
\cline{2-6}
$(N/Z)_{free}$ & $^{58}$Ni+$^{64}$Ni & 1.099 & 0.104 & 1.152 & 0.025\\
\cline{2-6}
& $^{64}$Ni+$^{64}$Ni & 1.351 & 0.124 & 1.311 & 0.031 \\
\hline
& $^{58}$Ni+$^{58}$Ni & 46.156 & 0.764 & 21.246 & 2.176\\
\cline{2-6}
$Z_{max}$ & $^{58}$Ni+$^{64}$Ni & 46.902 & 0.712 & 21.246 & 2.176\\
\cline{2-6}
& $^{64}$Ni+$^{64}$Ni & 47.770 & 0.671 & 22.889 & 1.991 \\
\hline
& $^{58}$Ni+$^{58}$Ni & 97.541 & 1.225 & 46.332 & 4.698 \\
\cline{2-6}
$A_{max}$ & $^{58}$Ni+$^{64}$Ni & 102.981 & 1.113 & 49.256 & 4.384\\
\cline{2-6}
& $^{64}$Ni+$^{64}$Ni & 108.620 & 1.155 & 51.634 & 4.478 \\
\hline
& $^{58}$Ni+$^{58}$Ni & 15.066 & 0.662 & 12.688 & 0.356\\
\cline{2-6}
$E_k $ & $^{58}$Ni+$^{64}$Ni & 15.194 & 0.511 & 12.776 & 0.326\\
\cline{2-6}
(MeV/A) & $^{64}$Ni+$^{64}$Ni & 15.224 & 0.543 & 12.844 & 0.389\\
\hline
\end{tabular}
\end{center}
\caption{average values and standard deviations of the different observables at freeze-out time ($t$=150 fm/c) and asymptotic time ($t$=500 fm/c) calculated for Sly5 EOS.}
\label{table_observable_Sly5}
\end{table}
\indent
Further insights can be obtained from Figure \ref{Largest_and_second_largest_vs_time}, which shows the time variation of the  largest  and second largest cluster mass  with time for the different  systems.\\
\indent
It is observed that the average  mass of the second largest cluster are maximum at $t$=500 fm/c. The same is true if we consider the charge instead of the mass (not shown). This is a strong indicator of the end of the dynamics.\\
\indent
At further times, the mass and charge of the residue still slowly decrease. A part of the reason is that evaporation is a very slow process, but this is also partially due to the finite lifetime of nuclei in the semiclassical approach. For this reason, the further evolution cannot be considered as a physical evaporation and we will  take $t$=500 fm/c as asymptotic time. If this time is doubled, the absolute values of the mass and charge of the evaporation residues slightly decrease, but all the qualitative conclusions of this paper are unchanged.\\
\indent
The values of the different observables examined in this paper at the freeze-out time and at the asymptotic time are reported for the three systems in Table \ref{table_observable_Sly5}. These predictions will serve us as reference for the study of the symmetry energy dependence.
At the asymptotic time $t=500$ fm/c, the average kinetic energy falls below the expected value at zero temperature $e_{kin,0}\approx 14$ MeV/A, which should be associated to the zero point motion considering the average density for the evaporation residue $\langle\rho\rangle\approx 0.08$ fm$^{-3}$. This is a clear indication that the further emission cannot be identified with physical evaporation from a hot source, but it is rather a drawback of the semiclassical treatment.\\
\indent
We now turn to examine the sensitivity of the global observables to the different coefficients of the symmetry energy.
The isoscalar EoS parameters are fixed by the Sly5 parametrization, and the isovector ones are independently varied around an average value given by the same Sly5 functional.
\begin{figure}[!ht]
\begin{center}
\includegraphics[width=0.9\columnwidth]{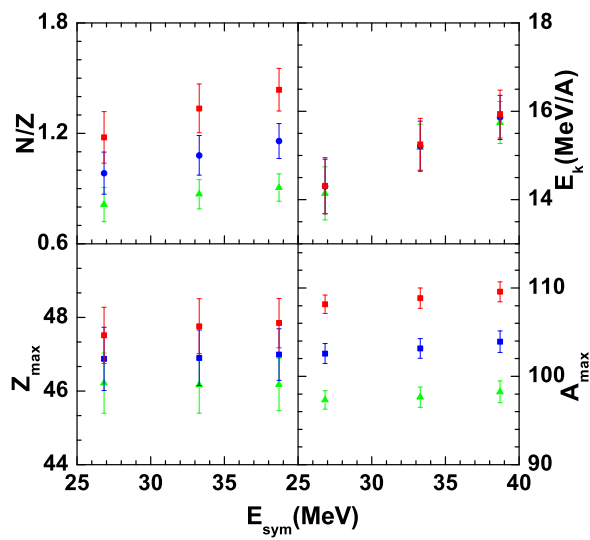}
\caption{Variation of $\langle N/Z \rangle$ of the unbound emitted particles (upper left), charge (lower left), mass (lower right) and total kinetic energy (upper right) of the fusion system  for independent variations of  $E_{sym}$, at the freeze-out time $t$=150 fm/c. The vertical error bars represent the standard deviation of the distributions.Results are given  for $^{58}Ni$ on $^{58}Ni$ (green triangles),  $^{58}Ni$ on $^{64}Ni$ (blue squares), $^{64}Ni$ on $^{64}Ni$ (red squares).}
\label{t150_esym}
\end{center}
\end{figure}
\begin{figure}[b]
\begin{center}
\includegraphics[width=0.9\columnwidth]{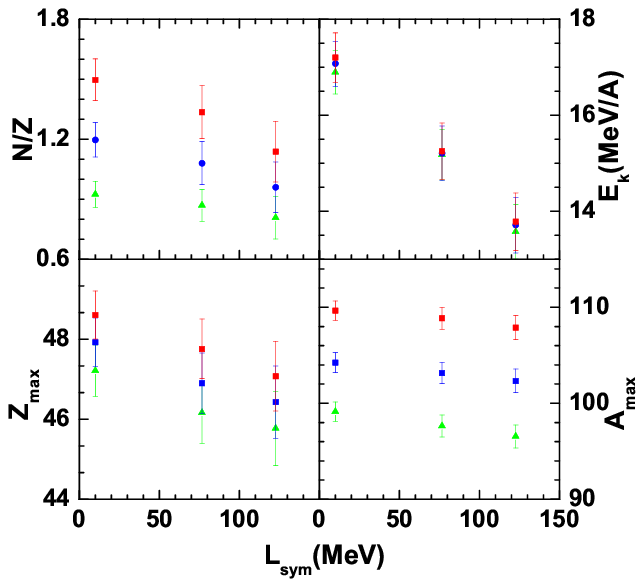}
\caption{ Variation of $\langle N/Z \rangle$ of the unbound emitted particles (upper left), charge (lower left), mass (lower right) and total kinetic energy (upper right) of the fusion system  for independent variations of  $L_{sym}$, at the freeze-out time $t$=150 fm/c. The vertical error bars represent the standard deviation of the distributions. Results are given  for $^{58}Ni$ on $^{58}Ni$ (green triangles),  $^{58}Ni$ on $^{64}Ni$ (blue squares), $^{64}Ni$ on $^{64}Ni$ (red squares).}
\label{t150_lsym}
\end{center}
\end{figure}

\begin{figure}[b]
\begin{center}
\includegraphics[width=0.9\columnwidth]{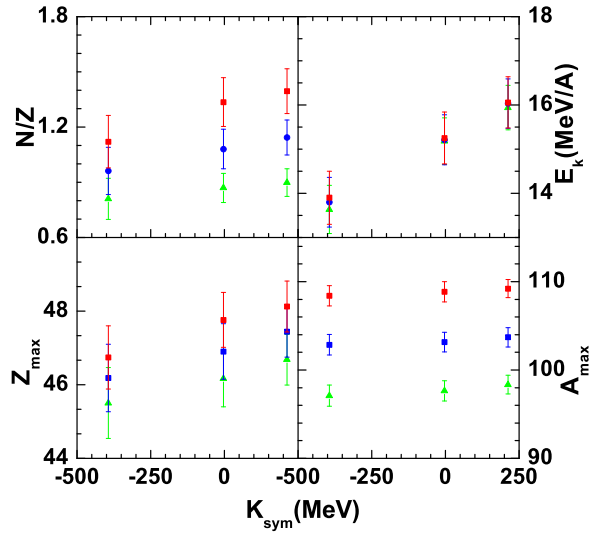}
\caption{ Variation of $\langle N/Z \rangle$ of the unbound emitted particles (upper left), charge (lower left), mass (lower right) and total kinetic energy (upper right) of the fusion system  for independent variations of   $K_{sym}$, at the freeze-out time $t$=150 fm/c. The vertical error bars represent the standard deviation of the distributions. Results are given  for $^{58}Ni$ on $^{58}Ni$ (green squares),  $^{58}Ni$ on $^{64}Ni$ (blue squares), $^{64}Ni$ on $^{64}Ni$ (red squares).}
\label{t150_ksym}
\end{center}
\end{figure}

\subsection{Sensitivity to the symmetry energy at freeze-out}\label{sec:freezeout}


This sensitivity study is first done at freeze-out time with respect to the three symmetry energy parameters $E_{sym}$, $L_{sym}$ and $K_{sym}$
in Figs. \ref{t150_esym},  \ref{t150_lsym} and \ref{t150_ksym} respectively.

Three values are considered for each parameter, within a variation domain
taken from ref.\cite{Margueron2018a}. In that work, those values are obtained from a compilation of present constraints extracted from empirical nuclear data. The  intermediate reference value for each parameter is given by the realistic functional Sly5, while the extreme minimum and maximum value cover the present uncertainties on the symmetry energy.
The values of the different parameters are reported in Table \ref{table_parameter}. The three parameters are independently varied,  that is two of them are kept fixed at the corresponding Sly5 value, while the third one is modified. Therefore, the point in each panel corresponding to the intermediate value of the symmetry parameter under study, represents the  predictions of the realistic EoS Sly5. These predictions are also reported in Table \ref{table_observable_Sly5}. The other points cannot be considered as realistic predictions, because physical correlations exist among the different empirical parameters, which are neglected here. However, they measure the sensitivity of the   observables to the different parameters describing the symmetry energy, and can therefore give information on the quantities that can be best constrained by intermediate energy heavy ion collisions.

As it is well known, pre-equilibrium neutrons are preferentially emitted by neutron-rich systems.
This very general feature is reproduced by our results.
However, the isospin excess is not entirely dissipated by pre-equilibrium emission.
Indeed we can observe that heavier and more neutron rich fusion sources are associated to heavier and more neutron rich systems. This appears to be a simple geometrical effect hardly related to the equation of state: the results are essentially unchanged if the parameters governing the density dependence of the symmetry energy are varied.

On the contrary, the pre-equilibrium emission is seen to be sensitive to the density dependence of the symmetry energy: higher values of $E_{sym}$  leading to more neutron rich emission. This finding is in good qualitative agreement with previous studies \cite{Bao-An,li_report} where the density dependence was controlled by a single parameter.
In our analysis, we can see that the effect of the different isospin parameters is very different:
an increasing $L_{sym}$ acts in the opposite direction as an increase in $E_{sym}$, while $K_{sym}$ is seen to play a similar role as $E_{sym}$. This can be understood from the fact that at these relatively low bombarding energies only subsaturation densities are explored. In the subsaturation density region, a higher value of symmetry energy slope ($L_{sym}$) leads to a lower symmetry energy whereas higher value of $E_{sym}$ and $K_{sym}$ represent higher symmetry energy. Similar behavior is also observed in the other observables shown in the next figures.

 The sensitivity of light particle emission to isospin is maximized at the earliest stage of the collision, as
it can be seen from Fig.\ref{N_by_Z_time_depndence}. This means that an amplified effect might be seen if kinematical cuts are employed to explicitly isolate the first chance emission \cite{li_report}. This analysis is left for future work.

Finally the upper right panel of  Figs. \ref{t150_esym},  \ref{t150_lsym} and \ref{t150_ksym} displays the average kinetic energy per nucleon of the residue  in its  reference frame. The internal kinetic energy is correlated to the excitation energy of the fused system.
We can see that this quantity is completely independent of the isospin content of the system. This is in agreement with the observation that the dynamical evolution is largely independent of the isospin, and a same freeze-out and asymptotic time can be associated to the three systems. It is interesting to remark that the density dependence of the symmetry energy is very influential in the determination of the kinetic energy, a stiffer EoS leading to a higher excitation.

\begin{figure}[!ht]
\begin{center}
\includegraphics[width=0.9\columnwidth]{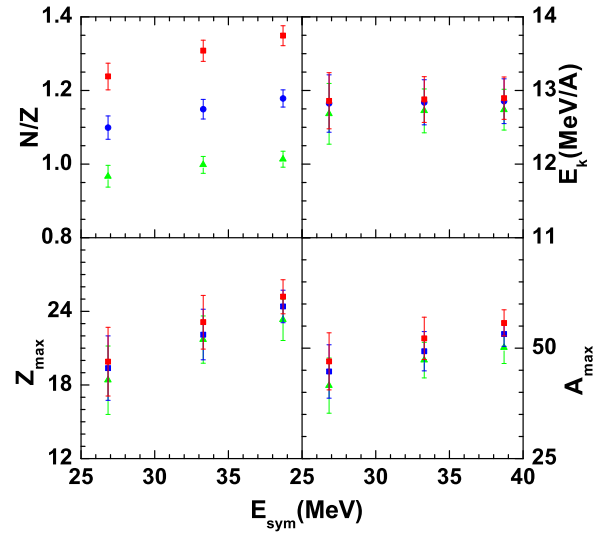}
\caption {Variation of $\langle N/Z \rangle$ of the unbound emitted particles (upper left), charge (lower left), mass (lower right) and total kinetic energy (upper right) of the fusion system  for independent variations of  $E_{sym}$, at the asymptotic time $t$=500 fm/c. The vertical error bars represent the standard deviation of the distributions. Results are given  for $^{58}Ni$ on $^{58}Ni$ (green triangles),  $^{58}Ni$ on $^{64}Ni$ (blue squares), $^{64}Ni$ on $^{64}Ni$ (red squares).}
\label{t500_esym}
\end{center}
\end{figure}

\begin{figure}[!ht]
\begin{center}
\includegraphics[width=0.9\columnwidth]{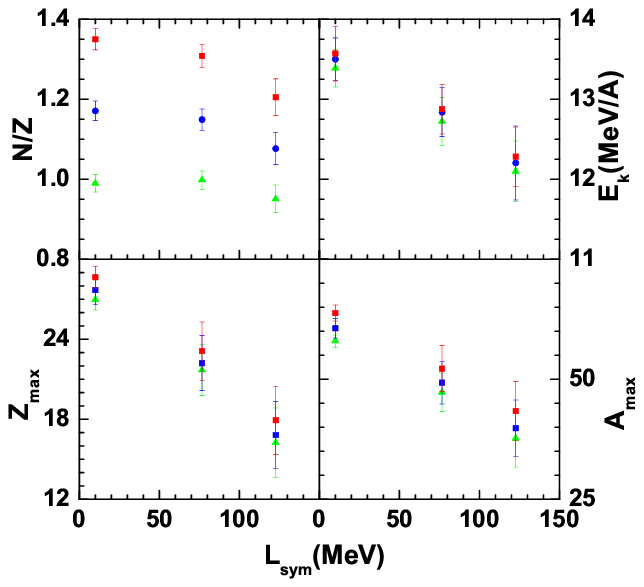}
\caption{ Variation of $\langle N/Z \rangle$ of the unbound emitted particles (upper left), charge (lower left), mass (lower right) and total kinetic energy (upper right) of the fusion system  for independent variations of  $L_{sym}$, at the   asymptotic time $t$=500 fm/c. The vertical error bars represent the standard deviation of the distributions. Results are given  for $^{58}Ni$ on $^{58}Ni$ (green triangles),  $^{58}Ni$ on $^{64}Ni$ (blue squares), $^{64}Ni$ on $^{64}Ni$ (red squares).}
\label{t500_lsym}
\end{center}
\end{figure}

\begin{figure}[!ht]
\begin{center}
\includegraphics[width=0.9\columnwidth]{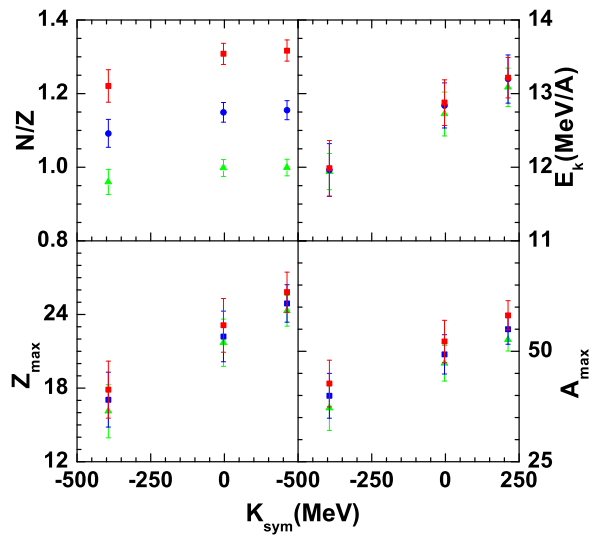}
\caption {Variation of $\langle N/Z \rangle$ of the unbound emitted particles (upper left), charge (lower left), mass (lower right) and total kinetic energy (upper right) of the fusion system  for independent variations of   $K_{sym}$, at the  asymptotic time $t$=500 fm/c. The vertical error bars represent the standard deviation of the distributions. Results are given  for $^{58}Ni$ on $^{58}Ni$ (green triangles),  $^{58}Ni$ on $^{64}Ni$ (blue squares), $^{64}Ni$ on $^{64}Ni$ (red squares).}
\label{t500_ksym}
\end{center}
\end{figure}

\subsection{Sensitivity to the symmetry energy at asymptotic times}


 We now turn to explore the sensitivity to the symmetry energy observed at the asymptotic stage of the reaction.
Indeed it was observed in previous studies using hybrid models that secondary decay can at least partially wash out the sensitivity to the equation of state \cite{Zhou,Mallik1,Ma}.

The same study done for Figs.\ref{t150_esym}, \ref{t150_lsym} and \ref{t150_ksym} is repeated at the asymptotic time
$t$=500 fm/c in Figs.\ref{t500_esym}, \ref{t500_lsym} and \ref{t500_ksym}. We can see that the effect of the secondary decay on the light particles is very small, as it could have been anticipated from Fig.\ref{N_by_Z_time_depndence}. As already observed above, this is most likely due to the fact that the characteristic pre-equilibrium time is smaller than the freeze-out time, and a deeper analysis is needed to deconvolute primary and secondary emitted particles. Still, the preserved sensitivity to the symmetry energy density dependence is encouraging because it suggests that the secondary decay should not blur up the signal.\\
\indent
 In all calculations, the kinetic energy at asymptotic time is reduced with respect to the values at freeze-out, due to particle evaporation. The residual internal kinetic energy is consistent with zero temperature Fermi motion at the typical average densities of the residues, $\rho\approx\rho_0/2$, consistent with the choice of asymptotic time discussed in Section \ref{sec:times}.
\\ \indent
The entrance channel dependence of the mass and charge of the heavy residue
appears reduced at the asymptotic time. This is however essentially a visul effect due to the different scales of the figures.
If we consider as a reference the calculation corresponding to Sly5 (see Table  \ref{table_observable_Sly5}), we can see that the average number of evaporated particles per nucleon is $A_{evap}/A^f_{max}=(A^f_{max}-A^a_{max})/A^f_{max}=0.52$ (superscript \enquote{f} and \enquote{a} represent values at freeze-out and asymptotic time respectively), independent of the system, and only a slight difference is observed for the average number
of evaporated protons per nucleon ($Z_{evap}/A^f_{max}=(Z^f_{max}-Z^a_{max})/A^f_{max}$) which changes from 0.23 for the most neutron rich, to  0.26 for the most neutron poor. This is consistent with the
constant excitation energy per nucleon at freeze-out observed in  Section \ref{sec:freezeout}.
\\ \indent
\begin{figure}[t]
\begin{center}
\includegraphics[width=0.9\columnwidth]{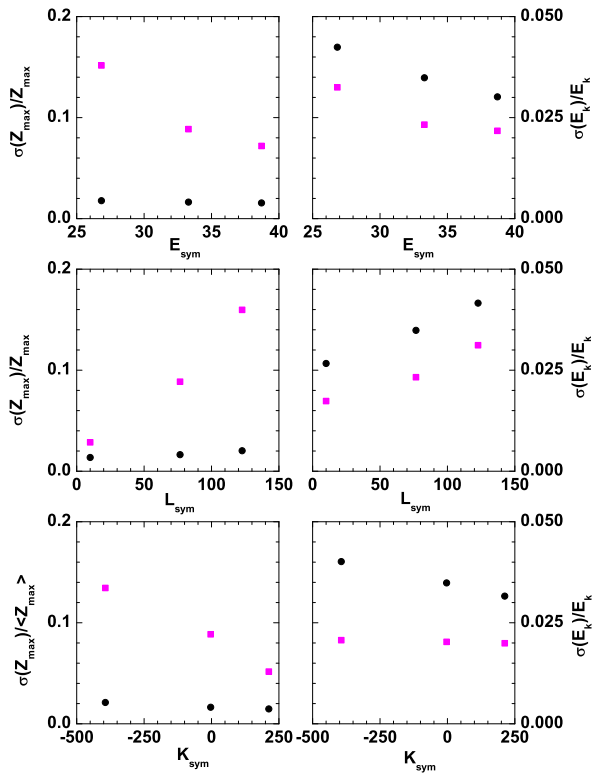}
\caption {Dependence of $\sigma(Z_{max})/Z_{max}$ (left panels) and $\sigma(E_{k})/E_{k}$ (right panels) of the heavy residue with $E_{sym}$ (upper panels), $L_{sym}$ (middle panels) and $K_{sym}$ (lower panels) for $^{58}Ni$ on $^{58}Ni$ reaction studied at freeze-out time $t$=150 fm/c (black circles) and asymptotic time $t$=500 fm/c (magenta squares).}
\label{variance_to_mean_ratio}
\end{center}
\end{figure}
Moreover, we can see from Figs.\ref{t500_esym}, \ref{t500_lsym} and \ref{t500_ksym} that
the sensitivity to the symmetry energy of $A_{max}$ and $Z_{max}$
is greatly amplified at the asymptotic time. The stiffer the equation of state at sub-saturation density,
the higher the mass and charge of the residue
\footnote{As already remarked above, a stiffer equation of state corresponds either to higher values of $E_{sym}$ and $K_{sym}$, or to lower values of $L_{sym}$.}.
This behavior seems in contradiction with the  behavior of the kinetic energy at freeze-out time observed in Figs.\ref{t150_esym},\ref{t150_lsym} and \ref{t150_ksym}.  The stiffer EoS being associated to a higher excitation at freeze-out, one would have expected lower mass and charge residues produced at the asymptotic times. A deeper analysis on the correlation between kinetic energy, excitation energy, and mass is needed to fully understand this point. This analysis is in progress, but some hints can already be obtained from inspection of Fig. \ref{variance_to_mean_ratio} .
\\
\indent
This figure displays the behavior of the variance to mean value ratio for the kinetic energy and charge of the heavy residue in the case of the $^{58}Ni$ on $^{58}Ni$ system. Both values at freeze-out and asymptotic  time are shown, as well as their dependence on $E_{sym}$, $L_{sym}$ and $K_{sym}$.
The other observables and the other entrance channels are not shown because they bear very similar information with respect to
Fig. \ref{variance_to_mean_ratio}.
We can see that at the freeze-out time the mass and charge fluctuations are very small and independent of the equation of state.
This is however not the case for the excitation energy.
Indeed, the highest average internal kinetic energies at freeze out, obtained from the stiffer equations of state, are systematically associated to the lowest relative dispersions. Because of  secondary evaporation, this sensitivity to the equation of state is transmitted to the asymptotic fluctuations of the mass and charge of the residue.
 \\
\indent
This ensemble of observations indicates that, in addition to the well-known sensitivity to the equation of state of pre-equilibrium particles, also  the size and charge of the evaporation residue can bring interesting information on the density dependence of the symmetry energy, and this is true both for their average values and their fluctuations. This sensitivity does not appear to be washed out by secondary decay, if the collision and the evaporation dynamics are consistently calculated using the same transport formalism.
In particular,  we observe a strong sensitivity to the $L_{sym}$ parameter which has been deeply studied in the astrophysical context, and shown to be well correlated to a number of astrophysical phenomena, such as the radius of neutron stars and the density of the crust-core transition.\\


 \section{Conclusions}
In this paper we have presented a first application of the BUU@VECC-McGill transport model including a realistic mean field functional with parameters optimized for the Sly5 effective interaction. The mean-field is implemented with a meta-modelling technique, that allows performing sensitivity studies of the measurable observables to the different
empirical parameters of the nuclear EoS. We have concentrated our study on the influence of the symmetry energy empirical parameters to the mass and charge of the fusion residues obtained in central Ni+Ni collisions with different isospin contents,  systems which are  going to be studied experimentally by the Indra/FAZIA collaboration in an upcoming experiment at Ganil.\\
\indent
To avoid the ambiguities in the definition of coupling time and coupling parameters with a statistical model, we have run the calculations up to an asymptotic time.
We have shown that the mass and charge of the residue are affected by the density dependence of the symmetry energy in an important way, while the isotopic ratio of free nucleons is less sensitive to the EoS. This can be understood from the fact that the free nucleon yield is dominated by the late stage of the collision, and kinematical cuts have to be imposed to recover the sensitivity to the EoS of the prompt nucleon emission. On the other side, energy conservation at freeze out imposes a lower average energy deposited in the fused system for a high symmetry energy below saturation (high $E_{sym}$ and $K_{sym}$ or low $L_{sym}$), and therefore a reduced effect on the secondary decay.\\
\indent
These observations suggest that the charge and isotopic composition of the fusion residue in intermediate energy $Ni+Ni$ heavy ion collisions can be an interesting probe of the symmetry energy.\\
\indent
An interesting aspect of the meta-modelling technique incorporated in the transport model is that the EoS extracted from microscopic ab-initio models can be directly implemented, thus reducing the uncertainty intervals presently existing on the different empirical parameters. This work is currently in progress.\\

\end{document}